\documentclass[prd,preprint,tightenlines,showpacs,preprintnumbers,nofootinbib,eqsecnum,superscriptaddress]{revtex4}

 \usepackage[dvips,final]{graphicx}
  \usepackage{amssymb}
   \usepackage{amsmath}
    \usepackage{amsfonts}
     \usepackage{bm}

\begin{document}

\title{Exclusive scalar glueball $f_0(1500)$ production \\
for FAIR and J-PARC energy range}

\author{A.~Szczurek}
\email{antoni.szczurek@ifj.edu.pl}
\affiliation{Institute of Nuclear Physics PAN, PL-31-342 Cracow,
Poland} 
\affiliation{University of Rzesz\'ow, PL-35-959 Rzesz\'ow, Poland}

\author{P.~Lebiedowicz}
\email{lebiedpiotr@o2.pl}
\affiliation{University of Rzesz\'ow, PL-35-959 Rzesz\'ow, Poland}





\date{\today}

\begin{abstract}
We evaluate differential distributions for exclusive scalar
$f_0(1500)$ production for $p \bar p  \to N_1 N_2 f_0$ (FAIR@GSI)
and $p p \to p p f_0$ (J-PARC@Tokai). Both QCD diffractive 
and pion-pion meson exchange current (MEC) components are included. 
Rather large cross sections are predicted.
The pion-pion component, never discussed in the literature, 
dominates close to the threshold 
while the diffractive component takes over for larger energies. 
The diffractive component is calculated based on 
two-gluon impact factors as well as in the framework 
of Khoze-Martin-Ryskin approach proposed for diffractive 
Higgs boson production.
Different unintegrated gluon distribution functions (UGDFs) from 
the literature are used. 
The production of $f_0(1500)$ close to threshold 
could limit the so-called $\pi NN$ form factor in the region of 
larger pion virtualities.
\end{abstract}


\maketitle

\section{Introduction}

Many theoretical calculations, including lattice QCD,
predicted existence of glueballs (particles dominantly made of 
gluons) with masses $M >$ 1.5 GeV. 
No one of them was up to now unambigously identified.
The lowest mass meson considered as a glueball candidate is 
a scalar $f_0(1500)$ \cite{AC96} discovered by the Crystall Barrel
Collaboration in proton-antiproton annihilation 
\cite{Crystall_Barrel_f0_1500}.
The branching fractions are consistent with the dominant glueball 
component \cite{anisovich}.
It was next observed by the WA102 collaboration in central
production in proton-proton collisions in two-pion
 \cite{WA102_f0_1500_2pi} and four-pion \cite{WA102_f0_1500_4pi}
decay channels at $\sqrt{s} \approx$ 30 GeV 
\footnote{No absolute normalization of the corresponding 
experimental cross section was available. 
Only two-pion or four-pion invariant mass spectra were discussed.}. 
Close and Kirk \cite{CK} proposed a phenomenological model
of central exclusive $f_0(1500)$ production. In their language
the pomerons (transverse and longitudinal) are the effective
(phenomenological) degrees of freedom \cite{CS99}.
The Close-Kirk amplitude was parametrized as
\begin{equation}
{\cal M}(t_1,t_2,\phi') = a_T \exp\left(\frac{b_T}{2}(t_1+t_2)\right) 
         + a_L \frac{\sqrt{t_1 t_2}}{\mu^2} 
          \exp\left(\frac{b_L}{2}(t_1+t_2)\right)\;\cos (\phi') \; .
\end{equation}
In their approach there is no explicit $f_0(1500)$-rapidity 
dependence of the corresponding amplitude.
Since the parameters were rather fitted to the not-normalized 
WA102 data \cite{WA102_f0_1500_2pi} no absolute normalization can 
be obtained within this approach. Furthermore the parametrization
is not giving energy dependence of the cross section,
so predictions for other (not-measured) energies are not possible.
In the present paper we will concentrate rather on a QCD-inspired 
approach.
It provides absolute normalization, energy dependence
and dependence on meson rapidity (or equivalently on $x_F$ of 
the meson).

The nature of the $f_0(1500)$ meson still remains rather 
unclear.
New large-scale devices being completed (J-PARC at Tokai) or 
planned in the future (FAIR at GSI) open a new possibility to study
the production of $f_0(1500)$ in more details.

In the present analysis we shall concentrate on exclusive
production of scalar $f_0(1500)$ in the following reactions:
\begin{equation}
\begin{split}
p + p &\to p + f_0(1500) + p \; , \\
p + \bar p &\to p + f_0(1500) + \bar p \; , \\
p + \bar p &\to n + f_0(1500) + \bar n \; , \\
\end{split}
\end{equation}
While the first process can be measured at J-PARC, the latter
two reactions could be measured by the PANDA Collaboration
at the new complex FAIR planned in GSI Darmstadt.
The combination of these processes could shed more light
on the mechanism of $f_0(1500)$ production as well as 
on its nature.

\begin{figure}[!htb]    
 \centerline{\includegraphics[width=0.4\textwidth]{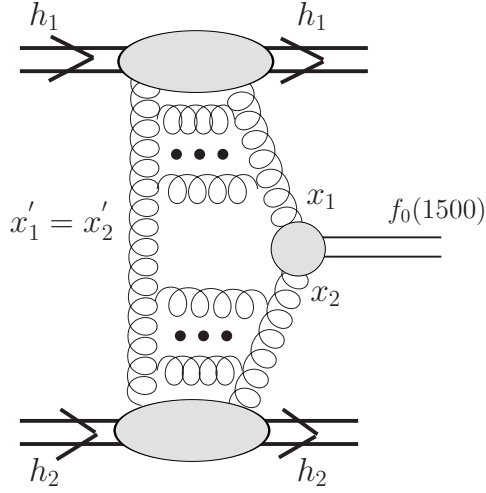}}
   \caption{\label{fig:QCDdiff}
   \small  The sketch of the bare QCD mechanism for diffractive
production of the glueball. The kinematical
variables are shown in addition.}
\end{figure}

If $f_0(1500)$ is a glueball (or has a strong glueball component \cite{CZ05})
then the mechanism shown in Fig. \ref{fig:QCDdiff}
may be important, at least in the high-energy regime.
This mechanism is often considered as the dominant mechanism of 
exclusive Higgs boson \cite{KMR}
and $\chi_c(0^+)$ meson \cite{PST07} production at high energies.
There is a hope to measure these processes at LHC in some future
when forward detectors will be completed.
At intermediate energies the same mechanism is, however, not able to
explain large cross section for exclusive $\eta'$ production 
\cite{SPT07} as measured by the WA102 collaboration. 
Explanation of this fact is not clear to us in the moment.

\begin{figure}[!htb]    
 \centerline{\includegraphics[width=0.3\textwidth]{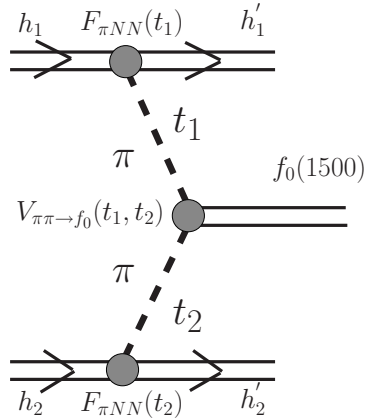}}
   \caption{\label{fig:pion_pion}
   \small  The sketch of the pion-pion MEC mechanism.
Form factors appearing in different vertices and kinematical variables
are shown explicitly.}
\end{figure}

At lower energies ($\sqrt{s} <$ 20 GeV) other processes may 
become important as well.
Since the two-pion channel is one of the dominant decay channels
of $f_0(1500)$ (34.9 $\pm$ 2.3 \%) \cite{PDG}
one may expect the two-pion fusion (see Fig.\ref{fig:pion_pion}
to be one of the dominant mechanisms of exclusive $f_0(1500)$ 
production at the FAIR energies. The two-pion fusion can be
also relative reliably calculated in the framework of 
meson exchange theory. The pion coupling to the nucleon is well
known \cite{Ericson-Weise}. The $\pi NN$ form factor for larger
pion virtualities is somewhat less known. This may limit our
predictions close to the threshold, where rather large 
virtualities are involved due to specific kinematics. 
At largest HESR (antiproton ring) energy, as will be discussed 
in the present paper, this is no longer a limiting factor 
as average pion virtualities are rather small. 

\section{Exclusive processes}

\subsection{Cross section and phase space}

The cross section for a general 3-body reaction $pp\to ppf_0(1500)$
can be written as
\begin{eqnarray}
d\sigma_{pp\to ppM}=\frac{1}{2\sqrt{s (s - 4 m^2)}}\,
\overline {|{\cal M}|^2} \cdot d^{\,3}PS \, . 
\label{general_cross_section_formula}
\end{eqnarray}
Above $m$ is the mass of the nucleon.

The three-body phase space volume element reads
\begin{eqnarray}
d^3 PS = \frac{d^3 p_1'}{2 E_1' (2 \pi)^3} \frac{d^3 p_2'}{2 E_2'
(2 \pi)^3} \frac{d^3 P_M}{2 E_M (2 \pi)^3} \cdot (2 \pi)^4
\delta^4 (p_1 + p_2 - p_1' - p_2' - P_M) \; . 
\label{dPS_element}
\end{eqnarray}
At high energies and small momentum transfers the phase space
volume element can be written as \cite{KMV99}
\begin{eqnarray}
d^3 PS \approx \frac{1}{2^8 \pi^4} dt_1 dt_2 d\xi_1 d\xi_2 d \phi \;
\delta \left( s(1-\xi_1)(1-\xi_2)-M^2 \right) \; ,
\label{dPS_element_he1}
\end{eqnarray}
where $\xi_1$, $\xi_2$ are longitudinal momentum fractions carried
by outgoing protons with respect to their parent protons and the
relative angle between outgoing protons $\Phi \in (0, 2\pi)$.
Changing variables  $(\xi_1, \xi_2) \to (x_F, M^2)$ one gets
\begin{eqnarray}
d^3 PS \approx \frac{1}{2^8 \pi^4} dt_1 dt_2 \frac{dx_F}{s \sqrt{x_F^2 +
4 (M^2+|{\bf P}_{M,t}|^2)/s}} \; d \Phi \; .
\label{dPS_element_he2}
\end{eqnarray}

The high-energy formulas (\ref{dPS_element_he1}) and
(\ref{dPS_element_he2}) break
close to the meson production threshold.
Then exact phase space formula (\ref{dPS_element}) must be taken 
and another choice of variables is more appropriate.
We choose transverse momenta of the outgoing nucleons ($p'_{1t}, p'_{2t}$),
azimuthal angle between outgoing nucleons ($\phi$) and rapidity of the
meson (y) as independent kinematically complete variables.
Then the cross section can be calculated as:
\begin{equation}
d \sigma = \sum_{k} {\cal J}^{-1}(p_{1t},p_{2t},\phi,y)|_{k} 
\frac{\overline{ |{\cal M}(p_{1t},p_{2t},\phi,y)|^2 }  }
{2 \sqrt{s (s-4 m^2)}} \; 
\frac{2 \pi}{(2 \pi)^5}
\frac{1}{2 E'_1} \frac{1}{2 E'_2} \frac{1}{2}
\;  p_{1t} p_{2t} d p_{1t} d p_{2t} d \phi dy \; ,
\end{equation}
where $k$ denotes symbolically discrete solutions of the set
of equations for $p'_{1z}$ and $p'_{2z}$:
\begin{eqnarray}
\left\{ \begin{array}{rcl}
\sqrt{s} - E_M  &=&
\sqrt{m_{1t}^2+p_{1z}^{'2}} + \sqrt{m_{2t}^2+p_{2z}^{'2}} \; ,  \\
-p_{Mz} &=& p'_{1z} + p'_{2z}   \; ,
\end{array} \right.
\label{energy_momentum_conservation}
\end{eqnarray}
where $m_{1t}$ and $m_{2t}$ are transverse masses of outgoing nucleons.
The solutions of Eq.(\ref{energy_momentum_conservation})
depend on the values of integration variables:
$p'_{1z} = p'_{1z}(p'_{1t},p'_{2t},\phi,y)$ and
$p'_{2z} = p'_{2z}(p'_{1t},p'_{2t},\phi,y)$.
The extra jacobian reads:
\begin{equation}
{\cal J}_k =
\left| \frac{p'_{1z}(k)}{\sqrt{m_{1t}^2+p'_{1z}(k)^2}} -
  \frac{p'_{2z}(k)}{\sqrt{m_{2t}^2+p'_{2z}(k)^2}} \right| \; .
\end{equation}
In the limit of high energies and central production, i.e. $p'_{1z} \gg$ 0 (very forward nucleon1), 
$-p'_{2z} \gg$ 0 (very backward nucleon2) the jacobian becomes a constant
${\cal J} \to \tfrac{1}{2}$.

The matrix element depends on the process and is a function of
kinematical variables. The mechanism of the exclusive production
of $f_0(1500)$ close to the threshold is not known. 
We shall address this issue here.
Therefore different mechanisms will be considered and the corresponding
cross sections will be calculated.

\subsection{Diffractive QCD amplitude}

According to Khoze-Martin-Ryskin approach (KMR) \cite{KMR}, we
write the amplitude of exclusive double diffractive colour singlet
production $pp\to ppf_0(1500)$ as
\begin{eqnarray}
{\cal
M}^{g^*g^*}=\frac{s}{2}\cdot\pi^2\frac12\frac{\delta_{c_1c_2}}{N_c^2-1}\,
\Im\int
d^2 q_{0,t}V^{c_1c_2}_J\frac{f^{off}_{g,1}(x_1,x_1',q_{0,t}^2,
q_{1,t}^2,t_1)f^{off}_{g,2}(x_2,x_2',q_{0,t}^2,q_{2,t}^2,t_2)}
{q_{0,t}^2\,q_{1,t}^2\, q_{2,t}^2}. \label{ampl}
\end{eqnarray}
The normalization of this amplitude differs from the KMR one
\cite{KMR} by the factor $s/2$ and coincides with the
normalization in our previous work on exclusive $\eta'$-production
\cite{SPT07}. The amplitude is averaged over the colour indices
and over two transverse polarisations of the incoming gluons
\cite{KMR}. The bare amplitude above is subjected to absorption
corrections which depend on collision energy
(the bigger the energy, the bigger the absorption corrections). 
We shall discuss this issue shortly when presenting our results.

The vertex factor $V_J^{c_1c_2}=V_J^{c_1c_2}(q_{1,t}^2,
q_{2,t}^2,P_{Mt}^2)$ in expression (\ref{ampl}) describes the
coupling of two virtual gluons to $f_0(1500)$
meson. Recently the vertex was obtained for off-shell values
of $q_{1,t}$ and $q_{2,t}$ in the case of $\chi_c(0)$ exclusive
production \cite{PST07}.
An almost alternative way to describe the vertex is to express it via
partial decay width $\Gamma(M \to gg)$. 
\footnote{The last value is not so well known. We shall take
$\Gamma(M \to gg) = \Gamma_M^{tot}$. This will give us un upper estimate.}
The latter (approximate) method can be used also for glueball production.

In the original Khoze-Martin-Ryskin (KMR) approach \cite{KMR}
the amplitude is written as
\begin{equation}
{\cal
  M}=N\int\frac{d^2q_{0,t}P[f_0(1500)]}{q_{0,t}^2q_{1,t}^2q_{2,t}^2}
f_g^{KMR}(x_1,x'_1,Q_{1,t}^2,\mu^2;t_1)f_g^{KMR}(x_2,x'_2,Q_{2,t}^2,\mu^2;t_2)
\; ,
\label{KMR_amplitude}
\end{equation}
where only one transverse momentum is taken into account somewhat
arbitrarily as
\begin{eqnarray}
Q_{1,t}^2=\mathrm{min}\{q_{0,t}^2,q_{1,t}^2\} \; , \qquad
Q_{2,t}^2=\mathrm{min}\{q_{0,t}^2,q_{2,t}^2\} \; ,
\label{glue_momenta}
\end{eqnarray}
and the normalization factor $N$ can be written in terms
of the $f_0(1500)\to gg$ decay width (see below).

In the KMR approach the large meson mass approximation
$M\gg |{\bf q}_{1,t}|,\,|{\bf q}_{2,t}|$ is adopted, so
the gluon virtualities are neglected in the vertex factor
\begin{equation}
P[f_0(1500)]\simeq(q_{1,t}q_{2,t})=(q_{0,t}+p'_{1,t})(q_{0,t}-p'_{2,t}).
\label{KMR_vert}
\end{equation}

The KMR UGDFs are written in the factorized form:
\begin{equation}
f_g^{KMR}(x,x',Q_t^2,\mu^2;t)=f_g^{KMR}(x,x',Q_t^2,\mu^2)\exp(b_0t)
\end{equation}
with $b_0=2$ GeV$^{-2}$ \cite{KMR}. In our approach we use
somewhat different parametrization of the $t$-dependent isoscalar form factors.
%

Please note that the KMR and our (general) skewed UGDFs have different
number of arguments. In the KMR approach there is only one effective
gluon transverse momentum (see Eq.(\ref{glue_momenta})) compared to two
idependent transverse momenta in general case (see Eq.(\ref{skewed_UGDFs})).

The KMR skewed distributions are given in terms of conventional
integrated densities $g$ and the so-called Sudakov form factor $T$ as
follows:
\begin{equation}
f_g^{KMR} (x,x',Q_t^2, \mu^2) = R_g
\frac{\partial}{\partial \ln Q_t^2}
\left[
\sqrt{T(Q_t^2,\mu^2)} x g(x,Q_t^2)
\right] \; .
\label{KMR_UGDF}
\end{equation}
The square root here was taken using arguments that only survival
probability for hard gluons is relevant.
It is not so-obvious if this approximation is reliable for light
meson production.
The factor $R_g$ in the KMR approach approximately accounts for 
the single $\log Q^2$ skewed effect \cite{KMR}. 
Please note also that in contrast to our approach
the skewed KMR UGDF does not explicitly depend on $x'$
(assuming $x' \ll x \ll 1$). Usually this factor is estimated to be
1.3--1.5. In our evaluations here we take it to be equal 1 to avoid
further uncertainties.
%
%
Following now the KMR notations we write the total
amplitude (\ref{ampl}) (averaged over colour and
polarisation states of incoming gluons) in the limit $M\gg
q_{1,t},\,q_{2,t}$ as
\begin{eqnarray}
{\cal M}=A\,\pi^2\,\frac{s}{2}\,\int d^2
q_{0,t}P[f_0(1500)]\frac{f^{off}_{g,1}(x_1,x_1',q_{0,t}^2,
q_{1,t}^2,t_1)f^{off}_{g,2}(x_2,x_2',q_{0,t}^2,q_{2,t}^2,t_2)}
{q_{0,t}^2\,q_{1,t}^2\, q_{2,t}^2}\,, \label{ampl-KMR}
\end{eqnarray}
where the normalization constant is
%
%

%
\begin{eqnarray}
A^2 = \frac{64\pi\Gamma(f_0(1500) \rightarrow
gg)}{(N_c^2-1)M^3} \; .
\label{A-from-Gamma}
\end{eqnarray}
%
%
%
%
%


In addition to the standard KMR approach we could use other
off-diagonal distributions (for details and a discussion see
\cite{SPT07,PST07}).
In the present work we shall use a few sets of unintegrated gluon
distributions which aim at the description of phenomena where
small gluon transverse momenta are involved. Some details
concerning the distributions can be found in Ref.~\cite{LS06}. We
shall follow the notation there.

In the general case we do not know off-diagonal UGDFs very well.
In \cite{SPT07,PST07} we have proposed a prescription how to calculate
the off-diagonal UGDFs:
\begin{eqnarray}\nonumber
f_{g,1}^{off} &=& \sqrt{f_{g}^{(1)}(x_1',q_{0,t}^2,\mu_0^2) \cdot
f_{g}^{(1)}(x_1,q_{1,t}^2,\mu^2)} \cdot F_1(t_1)\,, \\
f_{g,2}^{off} &=& \sqrt{f_{g}^{(2)}(x_2',q_{0,t}^2,\mu_0^2) \cdot
f_{g}^{(2)}(x_2,q_{2,t}^2,\mu^2)} \cdot F_1(t_2)\,,
\label{skewed_UGDFs}
\end{eqnarray}
where $F_1(t_1)$ and $F_1(t_2)$ are isoscalar nucleon form factors.
They can be parametrized as (\cite{PST07})
\begin{eqnarray}
F_1(t_{1,2}) = \frac{4 m_p^2 - 2.79\,t_{1,2}} {(4 m_p^2
-t_{1,2})(1-t_{1,2}/071)^2} \;.
\label{off-diag-formfactors}
\end{eqnarray}
Above $t_1$ and $t_2$ are total four-momentum transfers in the first
and second proton line, respectively.
While in the emission line the choice of the scale is rather
natural, there is no so-clear situation for the second screening-gluon
exchange
\cite{SPT07}.

Even at intermediate energies ($W$ = 10-50 GeV) typical 
$x_1^{'} = x_2^{'}$ are relatively small ($\sim$ 0.01).
However, characteristic $x_1, x_2 \sim M_{f_0}/\sqrt{s}$ are not 
too small (typically $>$ 10$^{-1}$). Therefore here we cannot use 
the small-$x$ models of UGDFs. In the latter case a Gaussian smearing
of the collinear distribution seems a reasonable solution:
\begin{eqnarray}
{\cal F}_{g}^{Gauss}(x,k_t^2,\mu_F^2)=xg^{coll}(x,\mu_F^2) \cdot
f_{Gauss}(k_t^2;\sigma_0)\;,
\label{Gaussian_UGDF}
\end{eqnarray}
where $g^{coll}(x,\mu_F^2)$ are standard collinear (integrated)
gluon distribution and $f_{Gauss}(k_t^2;\sigma_0)$ is a Gaussian
two-dimensional function
\begin{equation}
\begin{split}
f_{Gauss}(k_t^2,\sigma_0)=\frac{1}{2\pi\sigma_0^2} \exp\left(-k_t^2/2
\sigma_0^2\right)/\pi  \,. 
\label{Gaussian}
\end{split}
\end{equation}
Above $\sigma_0$ is a free parameter which one can expect to be of 
the order of 1 GeV. Based on our experience in \cite{SPT07} we expect
strong sensitivity to the actual value of the parameter $\sigma_0$.
Summarizing, a following prescription for the off-diagonal UGDF
seems reasonable:
\begin{equation}
f(x,x',k_t^2,k_t^{'2},t) = \sqrt{
f_{small-x}(x',k_t^{'2}) f_{Gauss}(x,k_t^2,\mu^2)
} \cdot F(t) \; ,
\label{mixed_prescription}
\end{equation}
where $f_{small-x}(x',k_t^{'2})$ is one of the typical small-$x$ 
UGDFs (see e.g.\cite{LS06}). So exemplary combinations are: 
KL $\otimes$ Gauss, BFKL $\otimes$ Gauss, GBW $\otimes$ Gauss 
(for notation see \cite{LS06}).
The natural choice of the scale is $\mu^2 = M_{f_0}^2$. 
This relatively low scale is possible with the
GRV-type of PDF parametrization \cite{GRV}.
We shall call (\ref{mixed_prescription}) a "mixed prescription"
for brevity.

\subsection{Two-gluon impact factor approach for subasymptotic energies}

The amplitude in the previous section, written in terms of 
off-diagonal UGDFs, was constructed for large energies.
The smaller the energy the shorter the QCD ladder. 
It is not obvious how to extrapolate the diffractive amplitude 
down to lower (close-to-threshold) energies.
Here we present slightly different method which seems more adequate
at lower energies.

At not too large energies the amplitude of elastic scattering
can be written as amplitude for two-gluon exchange 
\cite{pp_elastic_2g_if,SNS02}
\begin{equation}
{\cal M}_{pp \to pp}(s,t) =
i s \frac{N_c^2-1}{N_c^2}
\int d^2 k_t \; \alpha_s(k_{1t}^2) \alpha_s(k_{2t}^2)
\frac{ 3 F({\bf k}_{1t},{\bf k}_{2t})
       3 F({\bf k}_{1t},{\bf k}_{2t}) }
{(k_{1t}^2 + \mu_g^2) (k_{2t}^2 + \mu_g^2)} \; .
\label{2g_elastic_amplitude}
\end{equation}
In analogy to dipole-dipole or pion-pion scattering 
(see e.g. \cite{SNS02}) the impact factor can be parametrized as:
\begin{equation}
F({\bf k}_{1t},{\bf k}_{2t}) =
\frac{A^2}{A^2+({\bf k}_{1t} + {\bf k}_{2t})^2} -
\frac{A^2}{A^2+({\bf k}_{1t} - {\bf k}_{2t})^2} \; .
\label{impact_factor}
\end{equation}
At high energy the net four-momentum transfer:
$t = -({\bf k}_{1t}+{\bf k}_{2t})^2$.
$A$ in Eq.(\ref{2g_elastic_amplitude}) is a free parameter which
can be adjusted to elastic scattering. For our rough estimate
we take $A = m_{\rho}$.

\begin{figure}[!htb]    %
\includegraphics[width=0.35\textwidth]{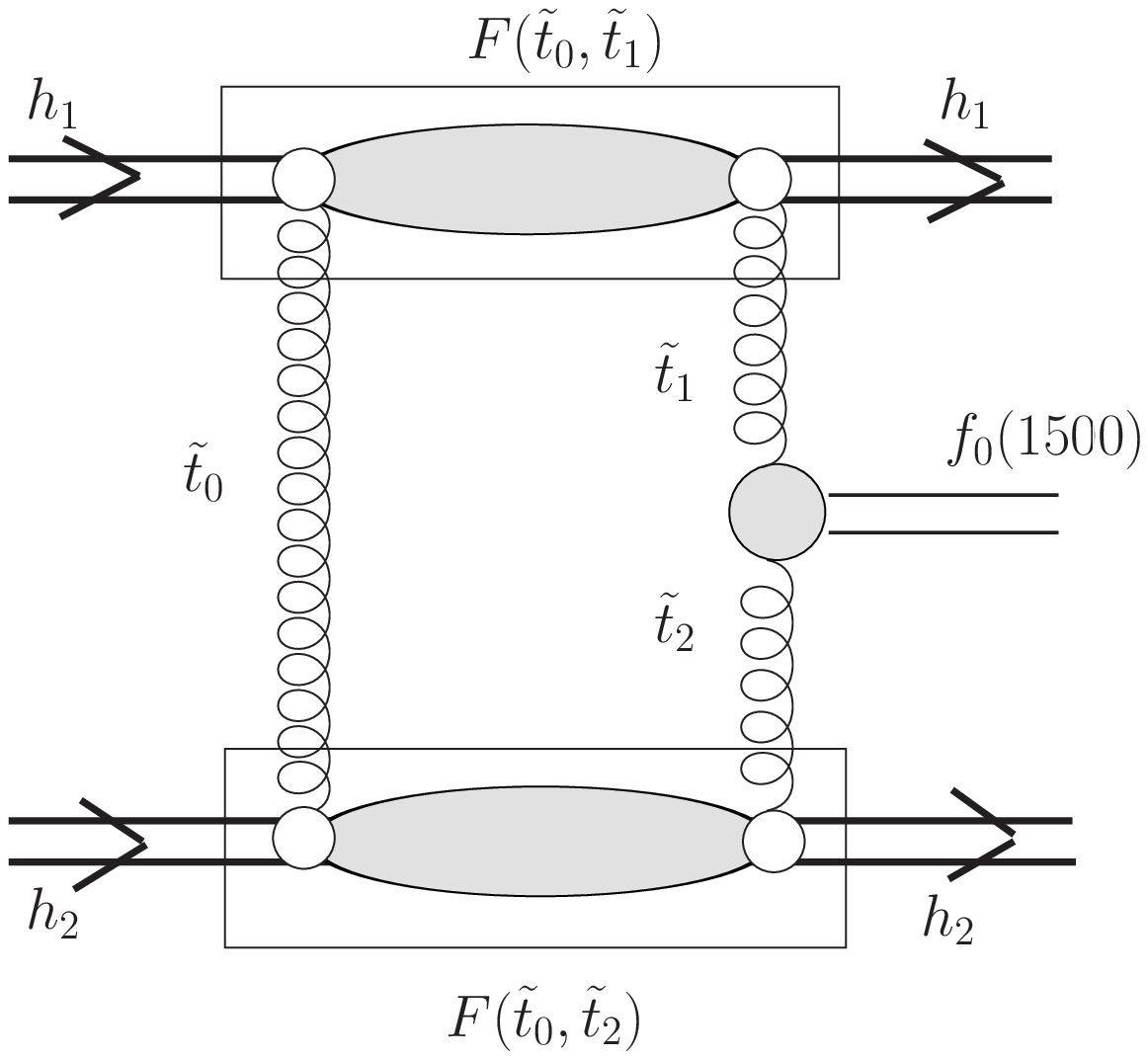}
\includegraphics[width=0.35\textwidth]{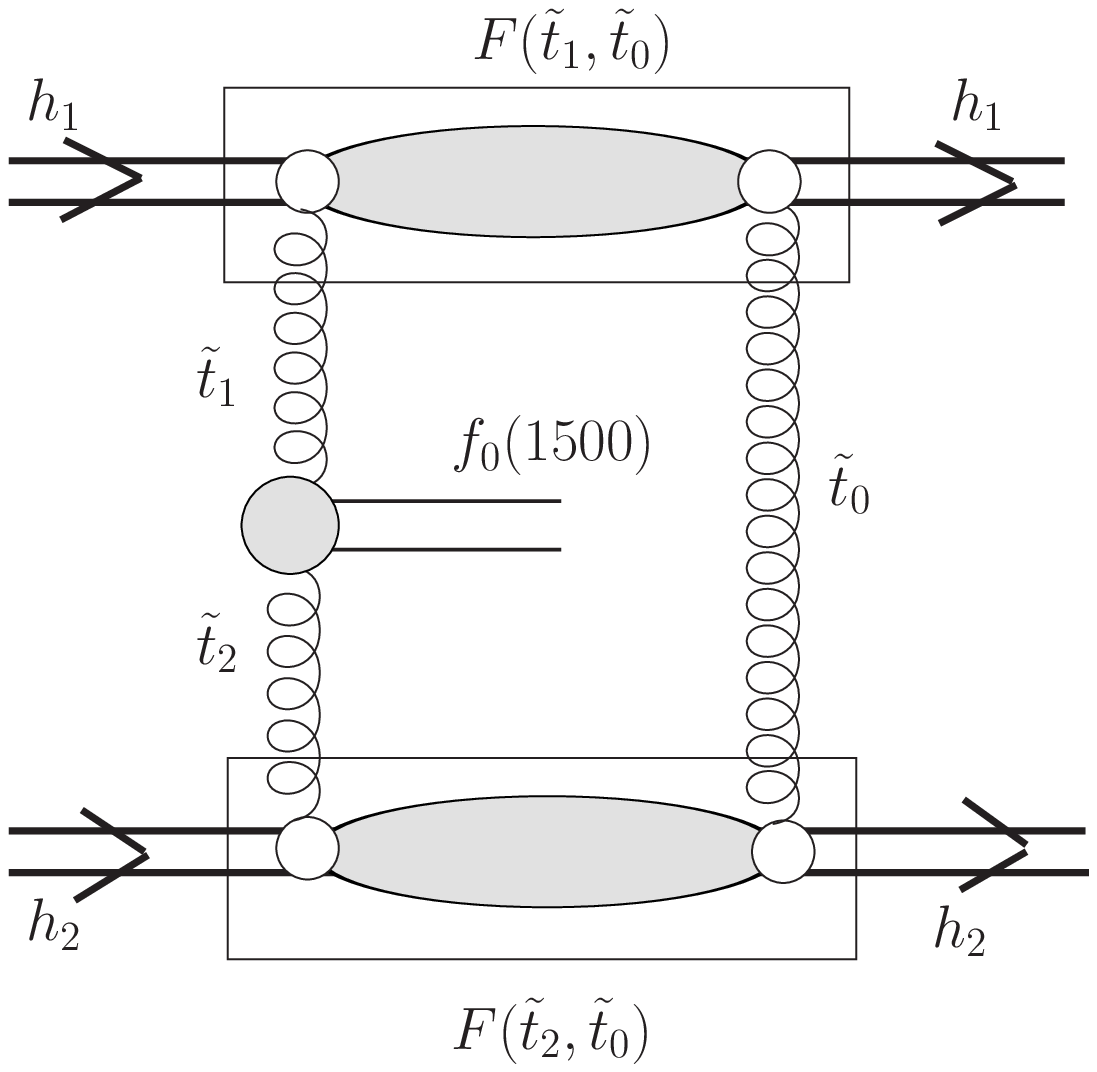}
   \caption{\label{fig:2g_impact_factor}
   \small  The sketch of the two-gluon impact factor approach.
Some kinematical variables are shown explicitly.}
\end{figure}

Generalizing, the amplitude for exclusive $f_0(1500)$ production
can be written as the amplitude for three-gluon exchange shown 
in Fig.\ref{fig:2g_impact_factor}:
\begin{eqnarray}
{\cal M}_{pp \to ppf_0(1500)}(s,y,t_1,t_2,\phi) =
i s \frac{N_c^2-1}{N_c^2}
\int d^2 k_{0t} \nonumber \\
\left( \alpha_s(k_{0t}^2) \alpha_s(k_{1t}^2) \right)^{1/2}
\left( \alpha_s(k_{0t}^2) \alpha_s(k_{2t}^2) \right)^{1/2} \nonumber \\
\frac{ 3  F({\bf k}_{0t},{\bf k}_{1t})
       3  F({\bf k}_{0t},{\bf k}_{2t}) }
{(k_{0t}^2 + \mu_g^2)
 (k_{1t}^2 + \mu_g^2)
 (k_{2t}^2 + \mu_g^2)} \;
V_{gg \to f_0(1500)}({\bf k}_{1t},{\bf k}_{2t})
 \; .
\label{3g_exclusive_amplitude}
\end{eqnarray}
At high energy and $y \approx$ 0 the four-momentum transfers can 
be calculated as:\\
$t_1 = -({\bf k}_{0t}+{\bf k}_{1t})^2$,
$t_2 = -({\bf k}_{0t}-{\bf k}_{2t})^2$. \\
At low energy and/or $y \ne$ 0 the kinematics is slightly more 
complicated.
Let us define effective four-vector transfers: 
\begin{eqnarray}
q_1 &=& (p'_1-p_1) = (q_{10},q_{1x},q_{1y},q_{1z}) \; ,  \nonumber \\
q_2 &=& (p'_2-p_2) = (q_{20},q_{2x},q_{2y},q_{2z}) \; .
\label{four-vector_tranfers}
\end{eqnarray}
Then $t_1 \equiv q_1^2 = q_{1l}^2 + q_{1t}^2$ and 
$t_2 \equiv q_2^2 = q_{2l}^2 + q_{2t}^2$.
Close to threshold the longitudinal components 
$q_{1l}^2 = q_{10}^2 - q_{1z}^2 \ll$ 0 and
$q_{2l}^2 = q_{20}^2 - q_{2z}^2 \ll$ 0.
Then the amplitude (\ref{3g_exclusive_amplitude}) must be corrected.
Then also four-vectors of exchanged gluons ($k_0$, $k_1$ and $k_2$)
cannot be purely transverse and longitudinal components must be included
as well.
To estimate the effect we use formula (\ref{3g_exclusive_amplitude})
\footnote{It would be more appropriate to calculate in this case
a four-dimensional integral instead of the two-dimensional one.}
but modify the transferred four momenta of gluons entering
the $g^* g^* \to f_0(1500)$ production vertex:
\begin{eqnarray}
k_1 &=& (0,{\bf k}_{1t},0) \to (q_{10},{\bf k}_{1t},q_{1z}) \; , \nonumber \\
k_2 &=& (0,{\bf k}_{2t},0) \to (q_{20},{\bf k}_{2t},q_{2z}) \; 
\label{4momenta_modification}
\end{eqnarray}
and leave $k_0$ purely transverse. This procedure is a bit arbitrary
but comparing results obtained with formula
(\ref{3g_exclusive_amplitude}) with that from the formula with modified
four-momenta would allow to estimate related uncertainties.

We write the vertex function
$g g \to f_0(1500)$ in the following tensorial form
\footnote{In general, another tensorial forms are also possible.
This may depend on the structure of the considered meson.}: 
\begin{equation}
V(k_1,k_2,p_M) = C_{f_0(1500) \to gg} \; g_{\mu \nu} k_1^{\mu} k_2^{\nu} \; .
\end{equation}
The normalization factor is obtained from the decay of $f_0(1500)$
into two soft gluons:
\begin{equation}
|C_{f_0(1500) \to gg}|^2 = \frac{64 \pi}{M_{f_0}^3(N_c^2-1)} 
\Gamma_{f_0(1500) \to g g} \; .
\label{vertex_normalization}
\end{equation}
Of course the partial decay width is limited from above:
\begin{equation}
\Gamma_{f_0(1500) \to gg} < \Gamma_{tot} \; .
\end{equation}

The amplitudes discussed here involve transverse momenta in the
infra-red region. Then a prescription how to extend the perturbative
$\alpha_s(k_t^2)$ dependence to a nonperturbative region
of small gluon virtualities is unavoidable. In the following
$\alpha_s(k_t^2)$ is obtained from an analytic freezing proposed by
Shirkov and Solovtsev \cite{SS97}.

\subsection{Pion-pion MEC amplitude}

It is straightforward to evaluate the pion-pion meson exchange
current contribution shown in Fig.\ref{fig:pion_pion}.
If we assume the $i \gamma_5$ type coupling of the pion to the nucleon 
then the Born amplitude reads:
\begin{eqnarray}
\overline{ |{\cal M}|^2} =
\frac{1}{4} 
&&[
\left( E_1 + m \right)
\left( E_1'+ m \right)
\left(
\frac{{\bf p}_1^2}{(E_1 + m)^2} + \frac{{\bf p}_1^{'2}}{(E_1^{'} + m)^2} -
\frac{ 2 {\bf p}_1 \cdot {\bf p}^{'}_1}{(E_1 + m)(E_1^{'} + m)}
\right)
] \cdot 2
\nonumber \\
&& 
\frac{g_{\pi NN}^2 \; \cdot T_k}{(t_1 - m_{\pi}^2)^2} F_{\pi NN}^2(t_1)
\; \cdot \; |C_{f_0(1500) \to \pi\pi}|^2
\; V_{\pi \pi \to f_0(1500)}^2(t_1,t_2) \; \cdot \;
\frac{g_{\pi NN}^2 \; \cdot T_k}{(t_2 - m_{\pi}^2)^2} F_{\pi NN}^2(t_2)
\nonumber \\
&&[
\left( E_2 + m \right)
\left( E_2'+ m \right)
\left(
\frac{{\bf p}_2^2}{(E_2 + m)^2} + \frac{{\bf p}_2^{'2}}{(E_2^{'} + m)^2} -
\frac{ 2 {\bf p}_2 \cdot {\bf p}^{'}_2}{(E_2 + m)(E_2^{'} + m)}
\right)
] \cdot 2 \; \nonumber \\.
\label{pion_pion_amplitude}
\end{eqnarray}
In the formula above $m$ is the mass of the nucleon,
$E_1, E_2$ and $E_1', E_2'$ are energies of initial and outgoing nucleons,
${\bf p}_1, {\bf p}_2$ and ${\bf p}'_1, {\bf p}'_2$ are corresponding
three-momenta and $m_{\pi}$ is the pion mass.
The factor $g_{\pi NN}$ is the familiar pion nucleon coupling constant
which is precisely known ($\frac{g_{\pi NN}^2}{4 \pi}$ = 13.5 -- 14.6).
The isospin factor $T_k$ equals 1 for the $\pi^0 \pi^0$ fusion
and equals 2 for the $\pi^+ \pi^-$ fusion. In the case of proton-proton
collisions only the $\pi^0 \pi^0$ fusion is allowed while in the case
of proton-antiproton collisions both $\pi^0 \pi^0$ and $\pi^+ \pi^-$
MEC are possible. In the case of central heavy meson production rather
large transverse momenta squared $t_1$ and $t_2$ are involved
and one has to include extended nature of the particles involved 
in corresponding vertices. This is incorporated via $F_{\pi NN}(t_1)$
or $F_{\pi NN}(t_2)$ vertex form factors. The influence of the
t-dependence of the form factors will be discussed in the result
section. In the meson exchange approach \cite{MHE87} they 
are parametrized in the monopole form as
\begin{equation}
F_{\pi N N}(t) = \frac{\Lambda^2 - m_{\pi}^2}{\Lambda^2 - t} \; .
\label{F_piNN_formfactor}
\end{equation} 
A typical values are $\Lambda$ = 1.2--1.4 GeV \cite{MHE87}.
The Gottfried Sum Rule violation prefers smaller $\Lambda \approx$
0.8 GeV \cite{GSR}.

The normalization constant $|C|^2$ in (\ref{pion_pion_amplitude})
can be calculated from the partial decay width as
\begin{equation}
|C_{f_0(1500) \to \pi \pi}|^2 
= \frac{8 \pi \; 2 M_{f_0}^2 \Gamma_{f_0(1500) \to \pi^0 \pi^0}}
{\sqrt{M_{f_0}^2 - 4 m_{\pi}^2 }}   \; ,
\label{normalization_constant}
\end{equation}
where $\Gamma_{f_0(1500) \to \pi^0 \pi^0} = 
0.109 \cdot  BR(f_0(1500) \to \pi \pi) \cdot 0.5$ GeV. The branching
ratio is $BR(f_0(1500) \to \pi \pi)$ = 0.349 \cite{PDG}.
The off-shellness of pions is also included for the 
$\pi \pi \to f_0(1500)$ transition through the extra 
$V_{\pi \pi \to f_0(1500)}(t_1,t_2)$ form factor 
which we take in the factorized form:
\begin{equation}
V_{\pi \pi \to f_0(1500)}(t_1,t_2) = 
\frac{\Lambda_{\pi\pi f_0}^2 - m_{\pi}^2}{\Lambda_{\pi\pi f_0}^2 - t_1} 
\cdot
\frac{\Lambda_{\pi\pi f_0}^2 - m_{\pi}^2}{\Lambda_{\pi\pi f_0}^2 - t_2}
\; .
\label{pipi2f0_formfactor}
\end{equation}
It is normalized to unity when both pions are on mass shell
\begin{equation}
V(t_1=m_{\pi}^2,t_2=m_{\pi}^2) = 1 \; . 
\end{equation}
In the present calculation we shall take $\Lambda_{\pi\pi f_0}$ = 1.0 GeV.

\section{Results}

\subsection{Gluonic QCD mechanisms}

Let us start with the QCD mechanism relevant at higher energies.
We wish to present differential distributions in 
$x_F$, $t_1$ or $t_2$ and relative azimuthal angle $\phi$.
In the following we shall assume: 
$\Gamma_{f_0(1500) \to gg} = \Gamma_{f_0(1500)}^{tot}$ .
This assumption means that our differential distributions
mean upper value of the cross section. If the fractional
branching ratio is known, our results should be multiplied
by its value.

\begin{figure}[!htb]    %
\includegraphics[width=0.55\textwidth]{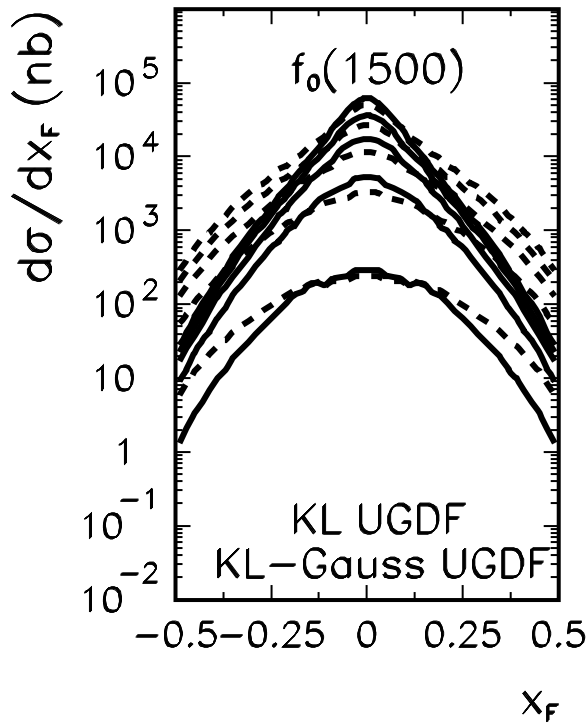}
   \caption{\label{fig:dsig_dxf_kl}
   \small  The distribution of $f_0(1500)$ in Feynman $x_F$ for 
   W = 10, 20, 30, 40, 50 GeV. In this calculation 
   the Kharzeev-Levin UGDF (solid line) and the mixed distribution 
   KL$\otimes$ Gauss (dashed line) were used.
}
\end{figure}

In Fig.\ref{fig:dsig_dxf_kl} we show as example
distribution in Feynman $x_F$ for Kharzeev-Levin UGDF (solid) and
the mixed distribution KL $\otimes$ Gaussian (dashed)
for several values of collision energy in the interval 
$W$ = 10 -- 50 GeV.
In general, the higher collision energy the larger cross section. 
With the rise of the initial energy the cross section becomes peaked 
more and more at $x_F \sim$ 0. The mixed UGDF produces
slightly broader distribution in $x_F$.


\begin{figure}[!htb]    %
\includegraphics[width=0.55\textwidth]{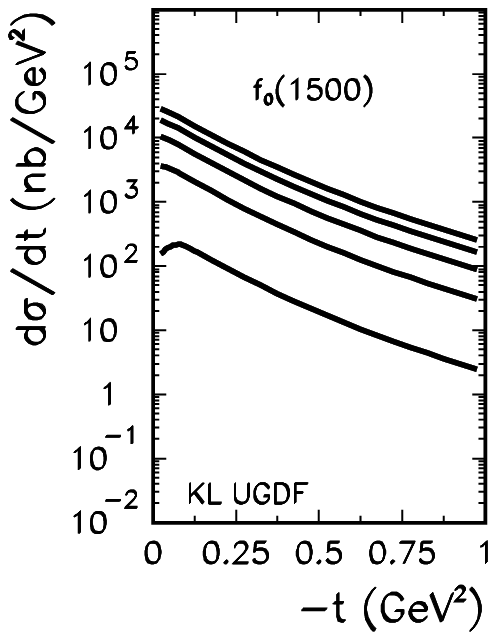}
   \caption{\label{fig:dsig_dt_kl}
   \small  Distribution in $t = t_1 = t_2$ for Kharzeev-Levin UGDF
for $W =$ 10, 20, 30, 40, 50 GeV.
The notation here is the same as in Fig.\ref{fig:dsig_dxf_kl}.
}
\end{figure}

In Fig.\ref{fig:dsig_dt_kl} we present corresponding distributions
in $t = t_1 = t_2$. The slope depends on UGDF used, but for a given
UGDF is almost energy independent.


Finally we present corresponding distributions in relative 
azimuthal angle between outgoing protons or proton and antiproton 
\footnote{The QCD gluonic mechanism is of course charge independent.}.
These distributions have maximum when outgoing nucleons are back-to-back.
Again the shape seems to be only weekly energy dependent.


\begin{figure}[!htb]    %
\includegraphics[width=0.55\textwidth]{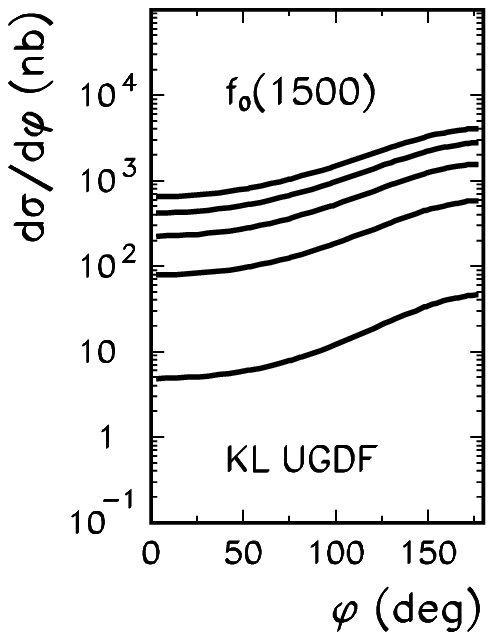}
   \caption{\label{fig:dsig_dphi_kl}
   \small  Distribution in relative azimuthal angle
for different UGDFs for $W$ = 10, 20, 30, 40, 50 GeV.
The notation here is the same as in Fig.\ref{fig:dsig_dxf_kl}.
}
\end{figure}


\subsection{Gluonic versus pion-pion mechanism}

\begin{figure}[!htb]    %
\includegraphics[width=0.45\textwidth]{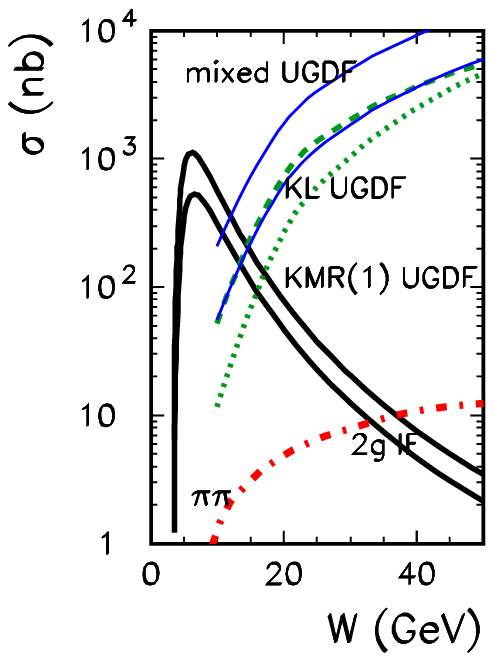}
\includegraphics[width=0.45\textwidth]{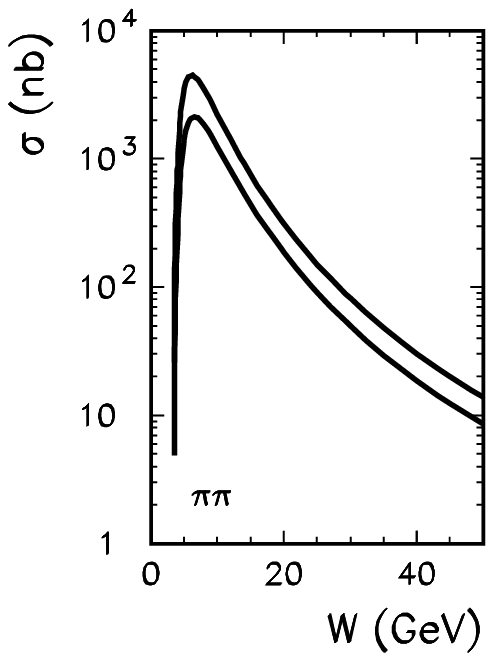}
   \caption{\label{fig:sigma_W}
   \small  The integrated cross section as a function
of the center of mass energy for $p \bar p \to p \bar p f_0(1500)$ 
(left panel)
and $p \bar p \to n \bar n f_0(1500)$ (right panel) reactions.
The thick solid lines are for pion-pion MEC contribution ($\Lambda$ = 0.8, 1.2 GeV), the dashed line is 
for QCD diffractive contribution obtained with the Kharzeev-Levin UGDF, 
the dotted line for the KMR approach
and the thin solid lines (blue on-line) are for "mixed" UGDF
(KL $\otimes$ Gaussian) with $\sigma_0$ = 0.5, 1 GeV. 
The dash-dotted line represents the two-gluon impact factor result.
}
\end{figure}

What about the pion-pion fusion mechanism? Can it dominate over
the gluonic mechanism discussed in the previous subsection?
In Fig.\ref{fig:sigma_W} we show the integrated cross section
for the exclusive $f_0(1500)$ elastic production
\begin{equation}
p \bar p \to p f_0(1500) \bar p 
\end{equation}
and for double charge exchange reaction
\begin{equation}
p \bar p \to n f_0(1500) \bar n  \; . 
\end{equation}
The thick solid line represents the pion-pion component calculated with 
monopole vertex form factors (\ref{F_piNN_formfactor}) with 
$\Lambda$ = 0.8 GeV (lower) and $\Lambda$ = 1.2 GeV (upper).
The difference between the lower and upper curves represents uncertainties
on the pion-pion compenent.
The pion-pion contribution grows quickly from the threshold, takes
maximum at $W \approx$ 6-7 GeV and then slowly drops with increasing
energy. The gluonic contribution calculated with unintegrated
gluon distributions drops with decreasing energy 
towards the kinematical threshold and seems to be about order of 
magnitude smaller than the pion-pion component at W = 10 GeV.
We show the result with Kharzeev-Levin UGDF (dashed line) which 
includes gluon saturation effects relevant for small-x, 
Kimber-Martin-Ryskin UGDF (dotted line) used for the exclusive 
production of the Higgs boson and the result with the 
"mixed prescription" (KL $\otimes$ Gaussian) for different
values of the $\sigma_0$ parameter: 0.5 GeV (upper thin solid line),
1.0 GeV (lower thin solid line). In the latter case results rather
strongly depend on the value of the smearing parameter.

We calculate the gluonic contribution down to W = 10 GeV.
Extrapolating the gluonic component to even lower energies in terms
of UGDFs seems rather unsure. At lower energies the two-gluon impact
factor approach seems more relevant.
The impact factor approach result is even order of magnitude smaller
than that calculated in the KMR approach (see lowest dash-dotted 
(red on-line) line in Fig. \ref{fig:sigma_W}), so it seems that 
the diffractive contribution is rather negligible at 
the FAIR energies.

Our calculation suggests that quite different energy dependence 
of the cross section may be expected in elastic and 
charge-exchange channels. Experimental studies at FAIR and J-PARC
could shed more light on the glueball production mechanism. 


\subsection{Predictions for PANDA at HESR}

Let us concentrate now on $p \bar p$ collisions at energies relevant
for future experiments at HESR at the FAIR facilite in GSI.
Here the pion-pion MEC (see Fig.\ref{fig:pion_pion}) seems 
to be the dominant mechanism, especially for the charge exchange 
reaction $ p \bar p \to n \bar n f_0(1500) $.

\begin{figure}[!htb]    %
\includegraphics[width=0.55\textwidth]{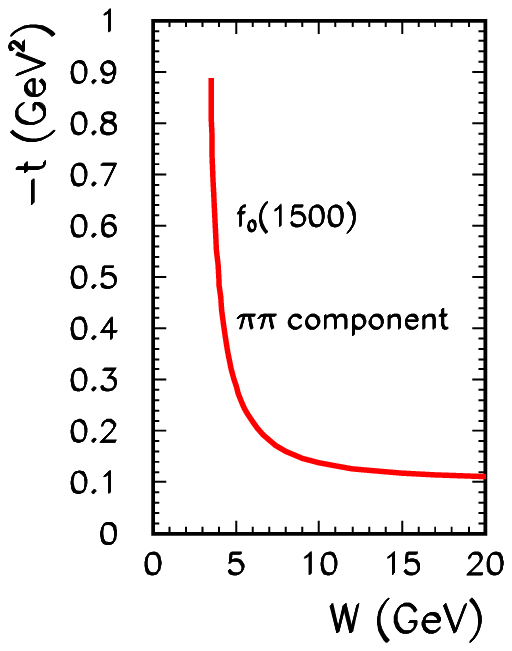}
   \caption{\label{fig:average_t1andt2_W}
   \small  
Average value of $ < t_1 > = < t_2> $ as a function of 
the center-of-mass collision energy for the two-pion exchange mechanism.
}
\end{figure}

In Fig.\ref{fig:average_t1andt2_W} we show average values
of $t_1$ (or $t_2$) for the two-pion MEC as a function of
the center of mass energy.
Close to threshold $W = 2 m_N + m_{f_0(1500)}$ the transferred
four-momenta squared are the biggest, of the order of about 
1.5 GeV$^2$.
The bigger energy the smaller the transferred four-momenta squared.
Therefore experiments close to threshold open a unique 
possibility to study physics of large transferred four-momenta
squared at relatively small energies.
This is a quite new region, which was not studied so far in 
the literature.


\begin{figure}[!htb]    %
\includegraphics[width=0.55\textwidth]{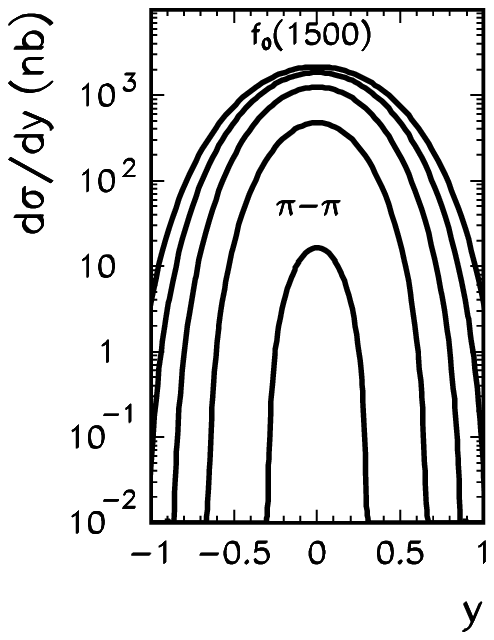}
   \caption{\label{fig:dsig_dy_pipi_f0_1500}
   \small  rapidity distribution of $f_0(1500)$
produced in the reaction $ p \bar p \to n \bar n f_0(1500) $
for W = 3.5, 4.0, 4.5, 5.0, 5.5 GeV (maximal HESR energy).
In this calculation $\Lambda$ = 0.8 GeV.
}
\end{figure}

The maximal energy planned for HESR is $\sqrt{s}$ = 5.5 GeV.
At this energy the phase space is still very limited.
In Fig.\ref{fig:dsig_dy_pipi_f0_1500} we show rapidity distribution
of $f_0(1500)$. For comparison the rapidity of incoming antiproton
and proton is 1.72 and -1.72, respectively. This means that in the
center-of-mass system the glueball is produced at midrapidities,
on average between rapidities of outgoing nucleons.


\begin{figure}[!htb]    %
\includegraphics[width=0.55\textwidth]{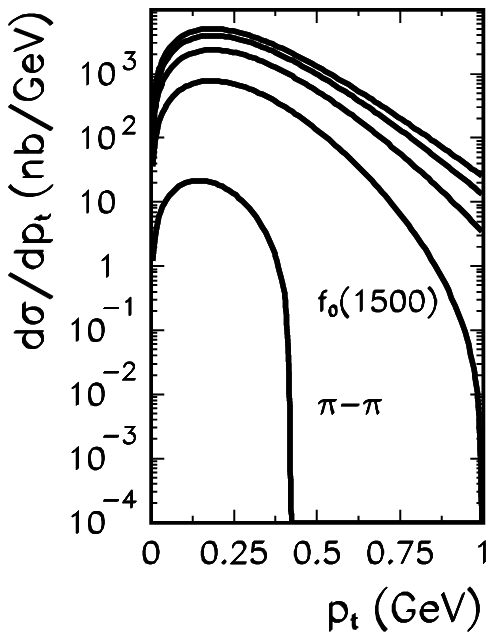}
   \caption{\label{fig:dsig_dpt_pipi_f0_1500}
   \small  Transverse momentum distribution of neutrons or antineutrons
produced in the reaction $ p \bar p \to n \bar n f_0(1500) $
for W = 3.5, 4.0, 4.5, 5.0, 5.5 GeV (maximal HESR energy).
In this calculation $\Lambda$ = 0.8 GeV.
}
\end{figure}

In Fig.\ref{fig:dsig_dpt_pipi_f0_1500} we show transverse
momentum distribution of neutrons or antineutrons produced in 
the reaction $ p \bar p \to n \bar n f_0(1500) $. 
The distribution depends on
the $\pi N N$ form factors $F_{\pi NN}(t_1)$ and $F_{\pi NN}(t_2)$ 
in formula (\ref{pion_pion_amplitude}). 


\begin{figure}[!htb]    %
\includegraphics[width=0.55\textwidth]{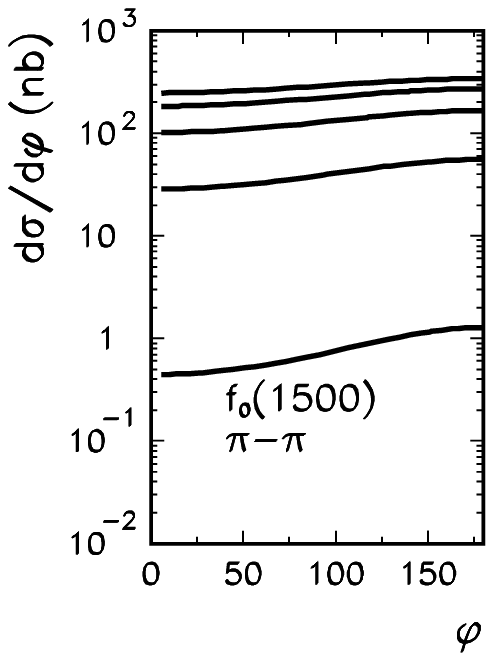}
   \caption{\label{fig:dsig_dphi_pipi_f0_1500}
   \small  Azimuthal angle correlations between neutron and antineutron
produced in the reaction $ p \bar p \to n \bar n f_0(1500) $
for W = 3.5, 4.0, 4.5, 5.0, 5.5 GeV (maximal HESR energy).
In this calculation $\Lambda$ = 0.8 GeV.
}
\end{figure}

In Fig.\ref{fig:dsig_dphi_pipi_f0_1500} we show azimuthal angle
correlation between outgoing hadrons (in this case neutron and
antineutron). The preference for back-to-back configurations is
caused merely by the limitations of the phase space close to 
the threshold.
This correlation vanishes in the limit of infinite energy.
In practice far from the threshold the distribution becomes almost 
constant in azimuth.  This has to be contrasted with similar 
distributions for pomeron-pomeron fusion shown in 
Fig.\ref{fig:dsig_dphi_kl}
which are clearly peaked for the back-to-back configurations.
Therefore a deviation from the constant distribution in relative 
azimuthal angle for the highest HESR energy of W = 5.5 GeV 
for $p \bar p \to p f_0(1500) \bar p$ can be a signal of the gluon 
induced processes. It is not well understood what happens with 
the gluon induced diffractive processes when going down to 
intermediate (W = 5-10 GeV) energies.
A future experiment performed by the PANDA collaboration could
bring new insights into this issue. This would be also a signal that
the $f_0(1500)$ state has a considerable glueball component.


\begin{figure}[!htb]    %
\includegraphics[width=0.45\textwidth]{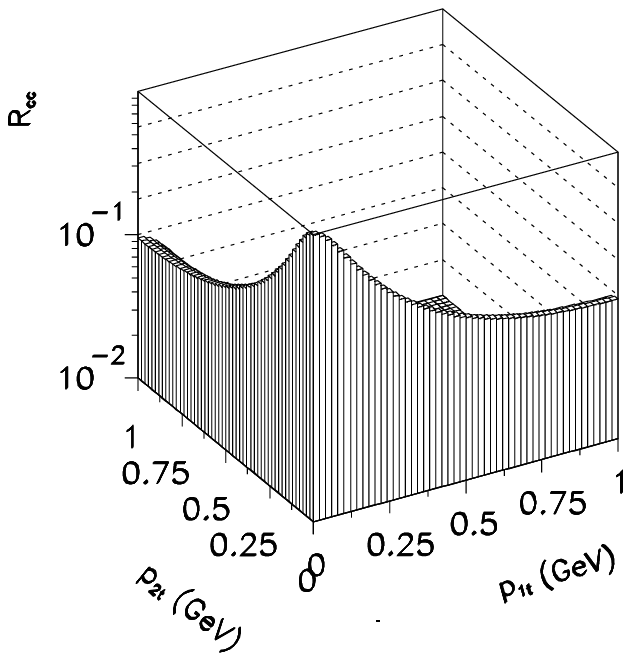}
\includegraphics[width=0.45\textwidth]{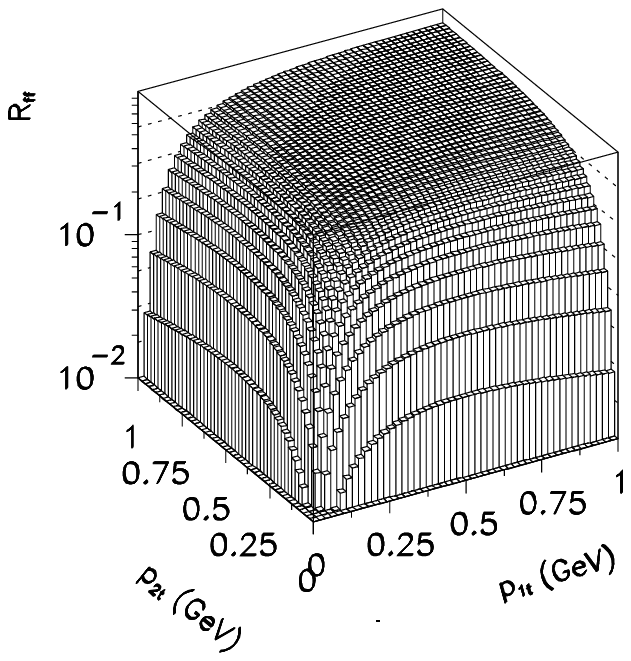}
\includegraphics[width=0.45\textwidth]{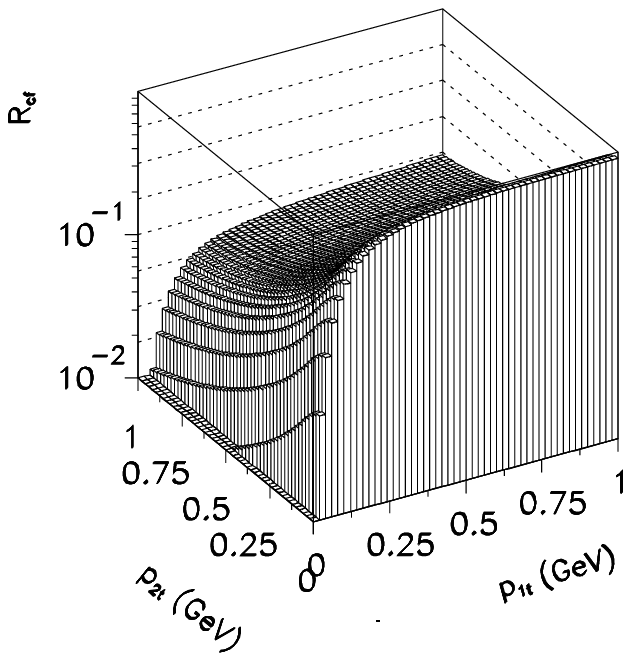}
\includegraphics[width=0.45\textwidth]{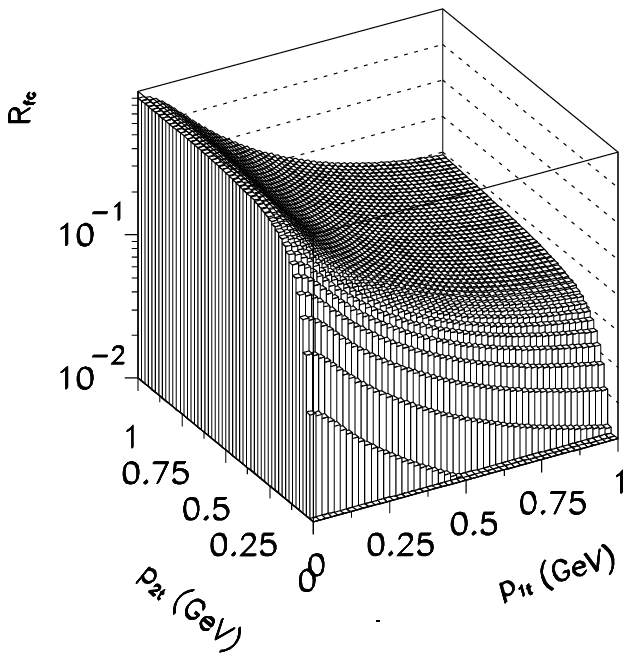}
   \caption{\label{fig:map_pt1pt2_pipi_f0_1500_rat}
   \small Helicity decomposition of the cross section
          on the ($p_{1t},p_{2t}$) plane for W = 10 GeV. 
          $R_{cc}$ (upper left), $R_{ff}$ (upper right),
          $R_{cf}$ (lower left), $R_{fc}$ (lower right).
          The standard nucleon dipole form factor was used in 
          this calculation.
}
\end{figure}

Up to now we have neglected interference between pion-pion and
pomeron-pomeron contributions (for the same final channel). 
This effect may be potentially important when both components are 
of the same order of magnitude.
While the pomeron-pomeron contribution is dominantly nucleon helicity
preserving the situation for pion-pion fusion is more complicated.
In the latter case we define 4 classes of contributions with respect
to the nucleon helicities: $cc$ (both helicity conserved),
$cf$ (first conserved, second flipped), $fc$ (first flipped, 
second conserved) and $ff$ (both helicities flipped).
The corresponding ratios of individual contributions to the sum of 
all contributions are shown in 
Fig.\ref{fig:map_pt1pt2_pipi_f0_1500_rat}. 
In practice, only the $cc$ $\pi \pi$ contribution may potentially 
interfere with the gluonic one. From the figure one can conclude 
that this can happen only when both transverse momenta of 
the final nucleons are small. We shall leave numerical studies of 
the interference effect for future investigations, when experimental 
details of such measurements will be better known; but already now one 
can expect them to be rather small.


\section{Discussion and Conclusions}

We have estimated the cross section for exclusive $f_0(1500)$ production
not far from the threshold. We have included both gluon induced diffractive
mechanism and the pion-pion exchange contributions. The first was obtained by
extrapolating down the cross section in the Khoze-Martin-Ryskin approach
with unintegrated gluon distributions from the literature as well as using
two-gluon impact factor approach. A rather large uncertainties are
associated with the diffractive component.
The calculation of MEC contribution requires
introducing extra vertex form factors which are not extremely well
constraint. This is especially important close to the threshold
where rather large $|t_1|$ and $|t_2|$ are involved.
The cross section for energies close to the threshold is very sensitive
to the functional form and parameters of vertex form factor.
Therefore a measurement of $f_0(1500)$ close to its production 
threshold could limit the so-called $\pi N N$ form factors 
in the region of exchanged four-momenta never tested before. 

We predict the dominance of the pion-pion contribution
close to the threshold and diffractive component far from the threshold.
Taking into account rather large uncertainties these predictions
should be taken with some grain of salt.
Clearly an experimental program is required to disantagle the reaction
mechanism.

Disantangling the mechanism of the exclusive $f_0(1500)$ production
not far from the meson production threshold would require study of
the $p \bar p \to p \bar p f_0(1500)$,
$p \bar p \to n \bar n f_0(1500)$ processes with PANDA detector 
at FAIR and $p p \to p p f_0(1500)$ reaction at J-PARC.
In the case the gluonic mechanisms are small and the pion exchange 
mechanism is a dominant process one expects:
$\sigma(p \bar p \to n \bar n f_0(1500)) = 4 \times
\sigma(p \bar p \to p \bar p f_0(1500))$.
On the other hand if the gluonic components dominate over MEC 
components
$\sigma(p \bar p \to p \bar p f_0(1500)) >
\sigma(p \bar p \to n \bar n f_0(1500))$.

Therefore a careful studies of different final channels at FAIR 
and J-PARC could help to shed light on coupling of (nonperturbative) 
gluons to $f_0(1500)$ and therefore would give a new hint on its 
nature.
Such studies are not easy at all as in the $\pi \pi$ decay channel one
expects a large continuum. This continuum requires a better 
theoretical estimate. A partial wave $\pi \pi$ analysis may 
be unavoidable in this context. The two-pion continuum will 
be studied in our future work.
A smaller continuum may be expected in the $K \bar K$ or 
four-pion $f_0(1500)$ decay channel. This requires, however, 
a good geometrical (full solid angle) coverage and high registration 
efficiencies.
PANDA detector seems to fullfil these requirements, but planning
real experiment requires a dedicated Monte Carlo simulation
of the apparatus.

\vskip 0.5cm

{\bf Acknowledgements} We are indebted to Roman Pasechnik, Wolfgang Sch\"afer
and Oleg Teryaev for a discussion and Tomasz Pietrycki for a help 
in preparing diagrams.


\end{document}